\documentclass[12pt,preprint]{aastex}

\begin{document}

\title{The Double-Lined Spectroscopic Binary Haro 1-14c}

\author{M. SIMON\altaffilmark{1} and L.PRATO\altaffilmark{2,3}}

\altaffiltext{1}{Department of Physics and Astronomy, SUNY,
Stony Brook, NY 11794-3800; michal.simon@sunysb.edu}

\altaffiltext{2}{Department of Physics and Astronomy, UCLA,
Los Angeles, CA 90095-1562; lprato@astro.ucla.edu}

\altaffiltext{3}{Current address: Lowell Observatory, 1400 West Mars Hill Road,
Flagstaff, AZ 86001}

\begin{abstract}
We report detection of the low-mass secondary in the spectroscopic binary
Haro 1-14c in the Ophiuchus star forming region.  The secondary/primary
mass ratio is $0.310\pm 0.014$.  With an estimated photometric primary mass of
1.2~$M_{\odot}$, the secondary mass is $\sim 0.4~M_{\odot}$ and
the projected semi-major axis is $\sim 1.5$ AU.  The system is well-suited 
for astrometric mapping of its orbit with the current generation of
ground-based IR interferometers.  This could yield precision values
of the system's  component masses and distance.  

\end{abstract}


\section{Introduction}

Reipurth et al. (2002) discovered that Haro 1-14c, a weak-lined
pre$-$main-sequence star
in the Ophiuchus star forming region (HBC 644, Herbig \& Bell
1988), is actually the primary component of a single-lined spectroscopic 
binary (SB1) with a period of 591 days.  We have been using high resolution 
infrared (IR) spectroscopy to detect the low-mass secondaries in main-sequence 
SB1s (Mazeh et al. 2003) and PMS SB1s (Prato et al. 2002a).
We included Haro 1-14c in our program for two 
reasons.  First, detection of its secondary would contribute to our study 
of the mass ratio distribution among PMS spectroscopic binaries
(e.g., Prato et al. 2002a, 2002b).  Second, it is important to test 
theoretical models of PMS evolution with stars whose masses have been
measured independently (e.g.,~Simon et al. 2000; Hillenbrand \& White 2004).
Meaningful tests require masses known to a few percent; this is possible
with masses measured dynamically.  The fairly long period of Haro 1-14c
and its location in the Ophiuchus SFR at $\sim 140$ pc distance suggested
that the orbit might be resolvable with the current generation interferometers
operating in the near IR.  Mapping observations of Haro 1-14c, combined 
with the parameters derivable from analysis of the binary as a double-lined 
spectroscopic binary (SB2), could yield precision measurements of
the primary and secondary masses and the
distance to the system (e.g., Boden et al. 2000). 
We have detected the secondary in Haro 1-14c in 6 observations distributed
over its orbital period.  This paper describes our results.

\section{Observations and Data Reduction}

We measured the radial velocities of the Haro 1-14c primary and secondary
at the Keck II telescope of the Keck Observatory using
the NIRSPEC high-resolution IR spectrometer
(McLean et al. 1998, 2000).  We centered the
spectrometer at 1.55 $\mu$m and obtained data following the procedures
we have described previously (Prato et al. 2002a,~2002b). All the 
observations were made without the adaptive optics system and have
average spectral resolution R $\sim$ 28,000.  The total integration times
ranged from 6$-$20 minutes, depending on seeing conditions, 
which provided spectra with signal-to-noise  ratio $\sim 100$.  The UT dates 
and modified Julian Dates (MJD) of the observations are listed in Table 1, 
columns 1 and 2. We used the REDSPEC software\footnote{See 
http://www2.keck.hawaii.edu/inst/nirspec/redspec/index.html} to extract
the spectra and place them on a wavelength scale.  As in our previous
work, we used only order 49 because it is almost completely free of
terrestrial absorption lines. 

To measure the radial velocities of the primary and secondary,
we used the TODCOR two-dimensional algorithm (Zucker \& Mazeh 1994)
and our suite of spectral templates as we have described before
(Mazeh et al. 2003, Prato et al. 2002a).  Of the templates available,
the main sequence star HR 8085, rotationally
broadened to 12 km s$^{-1}$, usually provided the best fit to the primary 
of Haro 1-14c.  Its K5 spectral type template is in reasonable agreement 
with the spectral type K3 listed in the Herbig and Bell Catalogue
(Herbig \& Bell 1988) for the component of Haro 1-14c 
brighter in visible light.  The M1.5 main sequence star GL 15A, also 
rotationally broadened to 12 km s$^{-1}$, fit the secondary the best.  
The average secondary/primary flux ratio at 1.55 $\mu$m was $\sim 0.4$.
However, because the main-sequence templates are not likely to match the
surface gravity and metallicity of the PMS components of Haro 1-14c exactly,
we do not regard as definitive the best-fitting spectral types and flux ratio. 
Table 1 lists our measured radial velocities of the primary  and secondary, 
$V_1$, and $V_2$, and their estimated uncertainties.

Reipurth et al.'s (2002) observations of Haro 1-14c provided 56 measurements
of its primary velocity and spanned $\sim 3.6$ orbital periods.  These yielded
accurate values of the binary center-of-mass velocity, $\gamma$, orbital
period, P, velocity semi-amplitude of the primary $K_1$, eccentricity, e, 
primary semi-major axis multiplied by the sine of the inclination, $a_1 sini$,
mass function, $f(M)$, time of periastron passage, $T_\circ$, and longitude
of the periastron of the primary, $\omega_1$.  To these parameters, our 
measurements of the system contribute uniquely the velocity semi-amplitude of 
the secondary, $K_2$, and hence the mass ratio, $q = M_2/M_1$, and
secondary semi-major axis, $a_2 sini$. We merged our
measurements with Reipurth et al's and, by least squares minimization of
a phase shift of the SB2 data, first improved the precision of the P and 
$T_\circ$ values enabled by the longer time span of the combined data.  
We then derived the other orbital parameters by least squares minimization;
Table 2 lists the values and Figure 1 shows the corresponding double-lined 
orbital solution.  The parameters in common with those calculated by Reipurth 
et al. are very close to their values, as expected, because the data set is 
dominated by their 56 measurements of the primary velocity.   The improvement 
in precision is greatest for $P$ and $T_\circ$ because of the longer
time interval of the merged data.  We used the new values of
$P$ and $T_\circ$ to calculate the phases of the SB2 observations listed 
in the last column of Table 1.     

\section{Discussion}

The K3 spectral type of the primary (Reipurth et al. 2002) and PMS
evolutionary tracks (e.g., Baraffe et al. 1998) for a few million year-old
star suggest that $M_1 \sim 1.2$ $M_{\odot}$.   This value, with the 
measured mass ratio, indicate that $M_2 \sim 0.4 M_{\odot}$ .  
From Table 1 we have the estimate   
$a~sini = (a_1 + a_2) sin i = 229.0 \pm8.0$ Gm $\sim 1.5$ AU.  At the
140 pc assumed distance to the Ophiuchus SFR, the apparent angular
size of the projected semi-major axis would be $\sim 11$ mas.  

Spectroscopic observations of SB2s leave three parameters undetermined: 
the two describing the binary orientation with respect to our line of 
sight, the orbital inclination, $i$, and the position angle of the line of
nodes, and the distance to the binary.  These parameters can be 
determined by resolving the components of the SB2 and mapping the orbit 
(e.g., Boden et al. 2000).
The requirements for a binary to be resolvable with the current generation
of IR interferometers are very stringent.  The apparent angular extent of
the binary orbit must be small enough that it lies within the diffraction
limit of a single telescope of the interferometer, but large enough that
it is resolvable at the projected interferometer baselines.  Both components
of the binary must be bright enough in the near-IR, usually K-band,
that fringes can be detected, and bright enough in the visible for wavefront 
correction by adaptive optics.  Targets satisfying these requirements
are rare.

With K$\sim7.8$ mag (2MASS catalog) and V$\sim 12.3$ mag (Rydgren et al. 1976),
and an estimated projected semi-major axis of 11 mas, Haro 1-14c
is suitable study for study by the combined techniques of
IR interferometry and spectroscopy.  The measurements will yield
the component masses, system distance (hence total luminosity),
component flux ratios at the wavelengths of the interferometric
and spectroscopic observations (enabling apportionment of the
luminosity between the components), and estimates of the effective
temperature (from the best matching templates for the analysis of
the spectra).   These parameters, together with the reasonable
assumption of equal age of the components, will enable definitive
tests of theoretical calculations of PMS evolution at masses below
0.5 $M_{\odot}$ (e.g. Simon et al. 2000; Hillenbrand and White 2004).

\section{Conclusions}

We have detected the secondary in the PMS binary Haro 1-14c by high 
spectral resolution observations at $1.55\mu$m.  We find that:
  
1) The mass ratio of $\sim 0.31$ suggests that for an estimated photometric
mass of 1.2 $M_{\odot}$ for the primary, the secondary mass is $\sim 0.4
M_{\odot}$.

2) Assuming these estimated masses, the projected orbital semi-major axis 
is $a~sini \sim 1.5$ AU. At an assumed 140 pc distance to the Ophiuchus 
SFR, the binary is angularly resolvable with the current generation of 
interferometers operating in the near-IR.

3) Astrometric measurements that map the binary orbit, and continued  
spectroscopic observations that improve the precision of the mass ratio
offer the promise to measure the distance to the binary and the
masses of its components at the few percent level.

\section{Acknowledgements}

We thank the referee for a constructive comment that helped us improve
the discussion section. 
We are very grateful to Tsevi Mazeh and Shay Zucker for allowing us to use 
their TODCOR algorithm.  We thank the Keck observing assistants,  and the 
staff in Waimea, for their thorough support.
This research was supported in part by NSF Grant AST 02-05427.  The data
were obtained at the W.M. Keck Observatory which is operated as a scientific
partnership between the California Institute of Technology, the University
of California, and NASA.  The observatory was made possible by the generous
financial support of the W.M. Keck Foundation.  Our research made use of 
data products from the Two Micron All Sky Survey, which is a joint project 
of the University of Massachusetts and the Infrared Processing and Analysis
Center/California Institute of Technology, funded by the National Aeronautics 
and Space Administration and the National Science Foundation.  Our research 
also used the SIMBAD database, operated at CDS, Strasbourg, France.
The authors extend special thanks to those of Hawaiian ancestry on whose 
sacred mountain we are privileged to be guests.

\clearpage

\begin{figure}
\plotone{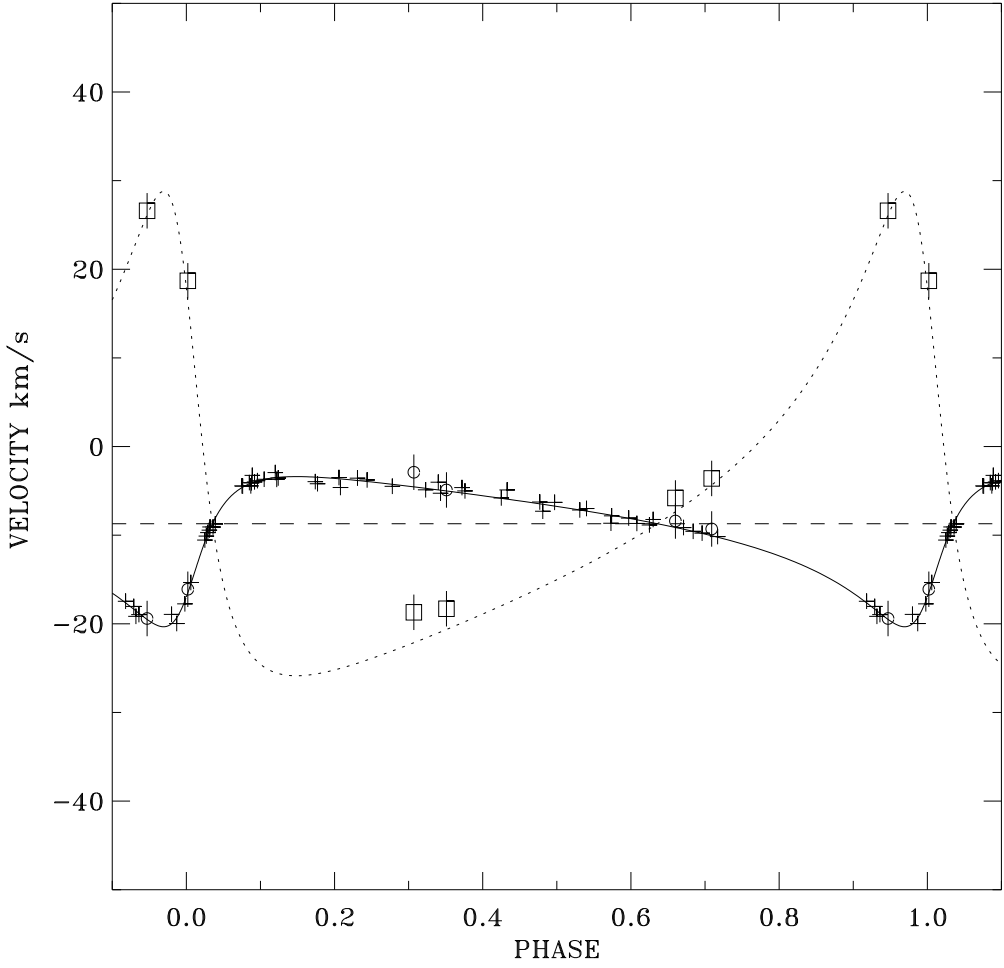}
\vskip 1.0cm
\caption{Radial velocity as a function of phase for Haro 1-14c.  The crosses
represent Reipurth et al.'s (2002) measurements of the primary velocity.
The open symbols, and 1 $\sigma$ uncertainties indicate our measurements
for the primary (circles) and secondary (squares).  The solid line shows 
the best fit to the primary star data, and the dashed line for the secondary.
The horizontal dashed line indicates the center-of-mass velocity of the
system.  The abscissa spans a 20\% redundancy in phase for clarity.
\label{fig1}}
\end{figure}   

\clearpage
\begin{deluxetable}{llccccc}
\tablewidth{0pt}
\tablecaption{Measured Velocities\label{tbl-1}}

\tablehead{
\colhead{UT Date} &\colhead{MJD} &\colhead{ $V_1$} &\colhead{$\sigma$}&
\colhead{$V_2$}&\colhead{$\sigma_2$}&\colhead{Phase}     \\
\colhead{} &\colhead{} &\colhead{km s$^{-1}$}&\colhead{km s$^{-1}$}&
\colhead{km s$^{-1}$} &\colhead{km s$^{-1}$} &\colhead{}}
\startdata
 2001 June 1  & 52061.50 &~ -2.9 & 2.0 & -18.7 & 2.0  & 0.307\\
 2002 July 17 & 52472.28 & -16.1 & 2.0  & ~18.7 & 2.0  & 0.002\\
 2003 Feb 8  &  52678.67 &~-4.9  & 2.0  &-18.3 & 2.0  & 0.351\\
 2003 Aug 10 &  52861.32 &~-8.4 & 2.0  &~~-5.8 & 2.0  & 0.660\\
 2003 Sep 8  &  52890.20 &~-9.3 & 2.0  &~~-3.6 & 2.0  & 0.709\\ 
 2004 Jan 26 &  53030.69 &-19.4 & 2.0  & ~26.6 & 2.0  & 0.947\\    
\enddata
\end{deluxetable}


\pagestyle{empty}

\begin{deluxetable}{ll}
\tablewidth{0pt}
\tablecaption{Orbital Elements\label{tbl-2}}
\tablehead{\colhead{Parameter}&\colhead{Value}}
\startdata
 $\gamma$                 &  $-8.71\pm0.07$  km s$^{-1}$ \\
 $P$                      &$591.3\pm0.3$ days     \\
 $K_1$                    &$8.52\pm0.13$ km s$^{-1}$   \\
 $K_2$                    &$27.5\pm 1.3$ km s$^{-1}$   \\
 $e$                      &$0.617\pm0.008$        \\  
 $q$                      &  $0.310\pm 0.014$     \\       
$a_1 sin i$               &$54.2\pm0.8$ Gm        \\
$a_2 sin i$               &$174.8\pm7.8$ Gm       \\
$M_1sin^3i$               &$0.98\pm0.14$ $M_{\odot}$   \\
$M_2sin^3i$               &$0.30\pm0.05$ $M_{\odot}$ \\ 
$\omega_1$                &$232.90\pm 0.55^\circ$  \\
 $T_\circ$                &$45375.4\pm3.0$ MJD \\

\enddata
\end{deluxetable}

\end{document}